%% file: efors.tex
\documentclass[]{aa}
\usepackage{graphicx}
\usepackage{txfonts}

\begin{document}

\title{ Astrometric precision of observations at VLT/FORS2}

\author{P.F.Lazorenko
	\inst{1} }

\offprints{P.F.Lazorenko}

\institute{Main Astronomical Observatory, National Academy of Sciencies of the
Ukraine, Zabolotnogo 27, 03680 Kyiv-127, Ukraine\\ email: laz@mao.kiev.ua  }

\date{}

\abstract{In this paper we test the  astrometric precision
of  VLT/FORS2 observations using
a serie of CCD frames taken in Galactic bulge area.
A special reduction method based on symmetrization of reference fields was used 
to reduce the atmospheric image motion.
Positional precision of  unsaturated  $R=16$~mag 
star images at 17~sec exposure and 0.55$\arcsec$ seeing 
was found to be equal to 300~$\mu$as. 
The total error of observations was decomposed into components.
It was shown  that  astrometric error  depends
largely on the photon centroiding error of the target (250~$\mu$as for
16~mag stars) while the image motion is much less (110~$\mu$as).
At galactic latitudes to about 20$\degr$, 
 precision  for a serie of frames with a 10~min total exposure
is estimated to be 30--50~$\mu$as for  14--16 mag stars providing the
images are not overexposed and the filter $R_{\rm special}$ is used. 
Error estimates for fields with smaller
sky star density are given. We conclude that astrometric observations
with large telescopes, under optimal reduction, are never atmospheric
limited.
The bias caused by differential chromatic refraction and residual
chromatism of LADC is considered and expressions  
valid for correcting color effects in the measured positions are given.

}


\maketitle

\section{Introduction}
The studies of exoplanets and  microlensing effects are usually restricted
to measuring  two physical quantities: radial velocities and/or brightness of
stars. Angular measurements are rarely considered due to their poor
precision. The extraction of many parameters related  to the studied 
object,  from a single physical quantity, 
however, is not always possible and reliable.
So, radial velocity data allow to find only the low limit of planet mass.
The additional information on the star angular
displacements essentially improves an accuracy of exoplanet mass
and orbital parameters determination (Pravdo \& Shaklan \cite{Pravdo};
Benedict et al. \cite{Benedict}). 
This  data are  also useful 
for microlensing studies of massive compact objects in the galaxy
because allow to measure lens mass, its distance and velocity 
(e.g. Boden et al. \cite{Boden}). Safizadeh et al. (\cite{Safizadeh})
argue that a combined use of photometric microlensing and astrometric
measurements allows detection of low mass planets, determination of their
masses and semi-major axes.

To be of practical importance, an accuracy of astrometric measurements
for the discussed purposes should be at least of the order of 10 to 100~$\mu$as.
A long history of ground-based observations testifies however 
that a real 
accuracy is much worse, about 1~mas per a night
for a 10--20$\arcmin$ reference frame  (Gatewood \cite{Gatewood})
and 1~mas/h for a 1$\arcmin$ double star
separation measurements (Han \cite{Han}). The best 150~$\mu$as/h precision
was achieved by Pravdo \& Shaklan (\cite{Pravdo}) in a special serie of 
observations at a 5-m Palomar telescope. A major factor limiting the accuracy
is the image motion caused by atmospheric turbulence and displayed as a 
random relative change of star image positions in unpredicted 
directions and at uncertain angle. Therefore development of ultra high-precision 
astrometric methods of observations from the ground  is
normally related to infrared interferometers which have
10~$\mu$as expected precision (e.g. Frink et al. \cite{Frink}).
A similar accuracy, at acceptable exposure times,  
was found to be unattainable for filled one-aperture telescopes 
due to   atmospheric image motion (Lindegren \cite{Lindegren}).

This conclusion was revised in the previous work 
(Lazorenko \& Lazorenko \cite{Lazorenko}, further 
Paper I) where  we considered the image motion as a turbulent light 
phase-related quantity dependent on initial wave-front fluctuations
at the telescope entrance pupil, on the process of differential measurements 
and on the way how these phase fluctuations
affect the  measured positions. In the spectral domain, 
transformation of phase fluctuations into image motion is described
by four filters of non-atmospheric origin which correspond to: 
1) conversion from the phase to the wave-front gradient,
2) averaging over the entrance pupil, 3) averaging due to a finite exposure and
4) formation of function differences in directions to reference stars.
Favorable for the inhibition of  atmospheric image motion spectrum
is that circumstance that a filter $Y(q)$ correspondent to the averaging over
the entrance pupil is a low-pass filter 
transparent to about $q \sim 2/D$ spatial frequency ($D$ is the telescope 
diameter) while reference field stars form a high-pass filter $Q(q)$.
The filter bands are only partially overlapping by their rising
(for $Q(q)$) and descending (for $Y(q)$) branches.
The shape of these instrumental functions can be adjusted so as to minimize
the combined system response $Y(q)Q(q)$. 
Improvement of the  $Y(q)$ filter shape
is achieved by apodization, or applying to the entrance pupil
a special covering with a variable light transmission. 
This measure results in a fast attenuation of the filter transmission at
$q > 2/D$.  The function $Q(q)$
shape is adjusted at the reduction phase by setting special weights to the
each reference star which virtually reconfigures the reference field
into a highly symmetric star group and supresses  
$Q(q)$ responce at low frequencies.

Attenuation factor of the  system with modified filters  $Y(q)$ and $Q(q)$
depends on the telescope diameter, angular reference field radius $R$,
the turbulent layer height $h$ and, being a power function of the ratio
$Rh/D$ is especially large for narrow-field mode of observations when 
$Rh \ll D$.
For that reason, the method is recommended for future large 30--100~m telescopes
though for existing 8--10~m instruments the expected precision is also high.
Based on theoretical asumptions, we had
shown that for 10-m  telescopes  precision of
differential referencing to the background stars
is, depending on the brightness and number 
of reference stars, from 10 to 60~$\mu$as at 10~min exposure. 

In the current study we test the  effeciency of this astrometric method 
application to actual observations at VLT and in the presence of
numerous noise sources, such as atmospheric image motion, photon noise in
the images of stars, pixelization, complex shape of PSF, background noise,
optical aberrations, active optics performance etc. 
In Sect.6 we discuss another atmospheric effect known as differential
chromatic   refraction (DCR) and which causes a relative displacement of
star images of different colors (Pravdo \& Shaklan \cite{Pravdo};
Monet et al. \cite {Monet}; Louarn et al. \cite{Louarn}).
At VLT, the DCR effect is strongly reduced by use of a special 
Longitudial Atmospheric Dispersion Compensator (LADC) which mimics atmospheric
refraction but introduces it in the opposite direction 
(Avila et al. \cite {Avila}). 

\section {Observational data and computation of centroids} 
For the test study we used a 4-hour serie of  FORS2 frames obtained
by Moutou et al. (\cite{Moutou}) when 
observing  OGLE-TR-132b object with the
filter $R_{\rm special}$, at  17~sec average exposure and a good 
FWHM=0.55$\arcsec $ seeing.
The  HR mode with $2 \times 2$ pixel binning and 0.125$\arcsec $ pixel size was
used. Only master chip of CCD 
with a field of view of $3.9 \times 2.1 \arcmin $ containing about
400 stars  brighter than 20 mag was considered, and only well exposed 
stars fainter than 16~mag($R$) were measured. Four frames obtained with $V$ filter
provided  color information used for examination of chromatic 
effects (Sect.6). 

A full profile fitting was used
for computation of  $ \bar {x}  $, 
$\bar {y}  $ coordinates of stellar centroids. The shape of PSF was
fitted by a 12 parameter model of a sum  of three elliptic Gaussians
with a common  $ \bar {x}  $, $ \bar {y}  $ centre: 
\begin{equation}
\label{eq:gauss}
\begin{array}{ll}
PSF(x,y)= & I G(x,y)+ (x-\bar{x})^2 I' G'(x,y) + \\
 & (y-\bar{y})^2 I'' G''(x,y)
\end{array}
\end{equation}
where $x $, $y $ are pixel coordinates.
The main Gaussian $G$ with 5 free parameters 
($ \bar {x}$, $\bar{y}$,
gaussian width parameters $\sigma_{0x}$, $\sigma_{0y}$,  a term $\alpha_0$
specifying orientation of semi-axes)  and a flux $I$ (sixth parameter) contains
about 98--99\% of the total star flux;
two auxiliary Gaussians  $G ' $ and $G''$ with fluxes $I ' $ and 
$I'' $ are oriented along  $x $, $y $ axes. As model parameters, they both
include $ \bar {x}$, $\bar{y}$,  two width terms $\sigma_{1x}$, $\sigma_{1y}$
for  $G ' $ and  similar $\sigma_{2x}$, $\sigma_{2y}$ terms for  $G''$.
Note that since image motion affects photocenter positions, it is useful
to treat $ \bar {x}$, $\bar{y}$ in  Eq.(\ref{eq:gauss}) not as the "fitting 
model center" but as the  {\it  weighted photocenter} defined by equations
$\int \int (x-\bar{x}) PSF(x,y) \, dx \, dy = 
\int \int (y-\bar{y}) PSF(x,y) \, dx \, dy =0$. This interpretation emphasizes
a necessity to use the PSF models with  first derivates on model parameters 
(except $ \bar {x}$, $\bar{y}$) being  symmetric functions
of coordinates $x$, $y$. 
In this case small errors in the parameter
determination induce symmetric bias of the PSF shape
which does not affect computation
of $ \bar {x}$, $\bar{y}$. For that reason, the model (\ref{eq:gauss}) does
not contain odd  coordinate powers.

Transition from the continuous  expression for the number of electrons
(\ref {eq:gauss})  to the  discrete pixel counts 
requires  numerical integration of  PSF$ (x, y) $ within the pixel limits.
To avoid this procedure, we used  analytical
approximation of the integral precise to  $ 10 ^ {-4} $.

Solution of Eq.(\ref{eq:gauss}) was obtained in square 10x10~px windows which
contained main light signal and are small enough to minimize overlapping 
of nearby images. 
Approximation of the actual PSF   shape by the model 
(\ref{eq:gauss}) typically is accurate to  5--10\%, the photon noise
statistics essentially differs from the Poissonian. 
Deviations of the actual PSF shape from the model have a specific
wavelike pattern which is highly correlated for all stars images of a certain
frame 
(especially at short spatial scales) 
but  completely changes at the next frame. 
At a high level of a light signal (bright images, central
pixels), these deviations strongly exceed random photon noise.

Using initial fitting for each star, we averaged
the "measured PSF - model" residuals over all star images in a frame
to find  systematic part of  these residuals as a function
var$PSF(x-\bar{x},y-\bar{y})$.
To obtain this function estimates at subpixel level, it was approximated
by a set
of 2D cubic polinomials  defined in $3 \times 3$~pixel areas 
with centres displaced 
at 1~pixel. The function var$PSF(x-\bar{x},y-\bar{y})$ was then subtracted
from individual star images and a final profile fitting applied.
This slightly improved  the precision of the centroiding.
One may note that the measured PSFs usually show variations at spatial
scales of some hundreds of pixels. Therefore it seems reasonable to obtain
more precise var$PSF(x-\bar{x},y-\bar{y})$ function shapes by performing
averaging of the "measured PSF - model" residuals only in the limited areas
of a frame. Leading to really better fitting of PSFs, this consideration
is however incorrect with regards to  determination of  photocenters
because application of variable 
image shape corrections dependent on position of a star in 
the frame induces a space-variable bias in  the weighted center positions. 

An attempt to assign weights to pixels depending on these residual values 
failed also,
though the quality of PSF approximation according to  $ \chi^2 $ 
criterion improved. This criterion seems to be improper for precise centroiding
works since inspite of low $\chi^2 \sim 10$  fitting quality for
bright stars, the actual centroiding precision is near the
photon limit (Section 4).

To estimate the centroiding precision for  images
with PSFs of  complex shape (\ref {eq:gauss}),
we created random images adding Poisson noise to the model function.
At 340~e$^-$/px sky level typical for
FORS2, the results are  represented by the expression
\begin{equation}
\label{eq:accur}
\varepsilon = \frac{ \mbox{FWHM}}{2.26\sqrt{I}}(1+ 475 I^{-0.7}) 
\end{equation}
Here FWHM refers to $x $ or $y $ axis and $I $ is given
in electrons. Note that Eq.(\ref {eq:accur}) is valid for random
photon fluctuations only and does not take into account 
a real photon statistics in central pixels. It was found  (Sect.4.1)
that  for bright images  a limit (\ref {eq:accur}) is not achievable.

\section{Reduction model} 
Astrometric reductions are based on theoretical considerations of Paper I which
main results are reproduced here in a concise form.

The variance $ \sigma_{\rm {at}} ^2 $ of differential
atmospheric image displacemets caused by a thin turbulent layer
can be found by integration of power spectrum of image motion 
which, as it was mentioned in Introduction, contains filter functions 
$Y(q)$ and $Q(q)$. Therefore,
using Eq.(12) of Paper I, we obtain
\begin{equation}
\label{eq:t1}
\sigma_{\rm{at}} ^2= \frac{J(h)}{VT} \int_{0}^{\infty} 
Y(q)
\left [ \sum_{m=1}^{\infty}q^{2m}F_{2m}(x_i,y_i) \right ] q^{-5/3} \, dq 
\end{equation}
where $T $ is the exposure,  $V $  is the  velocity of 
the turbulent layer and
$J(h) $ is a function proportional to the intensity of a 
turbulence in the layer and  its thickness.
The integration is performed in the area of a 2D spatial frequency $q $
related to the turbulent layer  plane. Cartesian star coordinates
$ x_i, y_i $ not distorted by the atmosphere, refer to the same plane.
The subintegral component in square brackets represents the $Q(q)$
function and describes filtration caused by  reference star  field. 
In original form (Eq.(25) of Paper I), it is written as 
a  linear combination of Bessel zero-order functions dependent on the star
coordinate differences. Note a modal structure of the $Q(q)$ component 
expansion into  $q ^ {2m} $  powers with coefficients $F_{2m} $ which are
functions  of $2m$ order cross-moments of  star  coordinates 
$ x_i, y_i $ differences measured with reference to the target.
The amplitude of  $q ^ {2m} F _ {2m} $ components rapidly  decreases
with $m $.
The $ Y(q)$ function describes
filtration  of the original image motion spectrum by the  
telescope entrance pupil which for better
inhibition of high-frequency spectral components 
can be apodized; for a filled pupil an expression for  $ Y(q)$ 
is given by Eq.(13) of Paper I. 

It is easy to formulate conditions at which 
the principal component $F_2 $ and possibly some
next  $F _ {2m} $ functions turn to zero at  a reduction stage.
For that purpose, it is necessary to define the target  position 
with reference to $N $  field  stars  by the quantity (Eq.(19) of Paper I)
\begin{equation}
\label{eq:t2}
W=  N^{-1} \sum_{i=1}a_i(\bar {x}_0-\bar {x}_i)=
	\bar {x}_0 - N^{-1} \sum_{i=1}a_i \bar {x}_i
\end{equation}
for the $x $-axis and by a similar quantity  for the $y $-axis.
Here $ \bar {x} _i $, $ \bar {y} _i $ are the measured (image motion included)
Cartesian coordinates of reference stars $i=1,2 \ldots N $; index $i=0 $ refers
to the target. Weights $ -\infty < a_i < \infty $ are determined from
conditions 
\begin {equation}
\label{eq:t3}
\begin{array}{lr}
\sum a_i (x_i-x_0)^{\alpha} (y_i-y_0)^{\beta}=0,& \;
		\alpha + \beta = 1 \ldots  \frac{k}{2}-1,   \\ 
\sum a_i =N,  \;\sum a_i^2 \varepsilon_i^2 = \mbox {min}   & \\
\end{array}
\end{equation}
where $\alpha$ and $\beta$ are non-negative integers, $\varepsilon_i$ is
the centroiding accuracy for $i$-th star. 
The first equation in (\ref{eq:t3}) describes
formally  a system of points with zero weighted cross-moments of the
$k/2-1 $  order and the symmetry  center  at $x_0 $, $y_0 $. The 
even parameter $k=2, 4 \ldots $  is called  therefore
the reference field symmetry order and is related 
to  the power of the first $q^k F^k$ non-zero component of Eq.(\ref {eq:t1}).
The last condition in (\ref {eq:t3}) is useful for minimization 
of reference field centroid errors
which affects the accuracy of  $W $ positions. 
The system (\ref {eq:t3}) is solved when at least $N _ {\rm {min}} =k
(k+2) /8 $ reference stars are available.

In the simplest case of $k=2 $ 
the first condition of the system (\ref {eq:t3})
is not used
and weights are found from equations 
$\sum a_i =N$, $\sum a_i^2 \varepsilon_i^2 = \mbox {min} $.
This refers, for example, to the double star ($N=1$) separation 
$W_x=(\bar {x}_0-\bar {x}_1)$ measurements or  to the 
average of individual separations $W_x=N^{-1} \sum (\bar {x}_0-\bar {x}_1)$
if $N>1$ reference stars of equal brightness are available.
The field is not symmetrized, and 
$ \sigma_{\rm {at}} ^2 $ value is maximum since the largest $q^2F_{2}$
term in Eq.(\ref{eq:t1}) is not eliminated. 

The first $k=4$ symmetrized solution is obtained applying conditions 
$\sum a_i (x_i-x_0)= \sum (y_i-y_0)=0$.
In particular, these conditions refer to the continuous distribution of
reference stars in a circular field
which, as shown by Lindegren (\cite{Lindegren}) yields a strong inhibition of
the image motion. The same $k=4$ symmetry order with elimination of
the  $q^2F_{2}$ term can be implemented with use of
only 3 arbitrary placed stars. The principal term in Eq.(\ref{eq:t1}) is
proportional to  $q^4F_{4}$.

The highest $k=12$ order implementation used in this study requires at least 
$N_{\rm {min}}=21$ reference stars to apply conditions
$\sum a_i (x_i-x_0)= \sum  a_i(y_i-y_0)=\sum a_i (x_i-x_0)^2= 
\sum  a_i(x_i-x_0)(y_i-y_0)= \ldots =  \sum a_i (x_i-x_0)(y_i-y_0)^4= 
\sum a_i (y_i-y_0)^5= 0$. This results in elimination of $F_2$, $F_4$, $F_6$, 
$F_8$ and  $F_{10}$ terms thus $Q(q)$ is proportional to $q^{12}F_{12}$. 

In proper motion works, the displacement  $ \mu_x $ of the target star
is found as a difference of $W (Ep2) -W (Ep1) $ at epochs Ep2 and Ep1. 
Weights $a_i $ are determined from conditions (\ref {eq:t3}) in which
the measured positions at Ep1 (standard frame) 
are used as unknown true coordinates 
$x_i =\bar{x} _i (Ep1) $, $y_i =\bar {y} _i (Ep1) $. 
This substitution,
as seen from Eqs.(\ref {eq:t3}), leads to the identity $W (Ep1)=0 $ 
that defines a conditional zero-point  of $W$ values.
Positions $W (Ep1)$ define also a zero-point of proper motions
which are computed as
\begin{equation}
\label{eq:m}
\begin{array}{l}
\mu_x= W(Ep2)= \frac{1}{N}\sum a_i (\bar{x}_0-\bar{x}_i)=
\bar{x}_0 - x_0 +  \frac{1}{N}\sum a_i (x_i-\bar{x}_i)
\end{array}
\end{equation}

Symmetrization of reference fields not only suppress 
atmospheric image motion, but also reduces geometrical  field
distortions, in this respect being  equivalent to usual
reduction models. Let's consider astrometric 
reduction model based on a polynomial  expansion
\begin{equation}
\label{eq:r}
\begin{array}{ll}
\bar{x}_i - x_i =&  A_0+A_1^{1,0}(x_i-x_0)+A_1^{0,1}(y_i-y_0)+ \ldots =\\
	& A_0 +\sum \limits_{m=1}^{k/2-1} \sum \limits_{ \begin{array}{c}
					^{\alpha,\beta =0,}
					_{\alpha+\beta = m }
					\end{array} }   ^m
	A_m^{\alpha,\beta }(x_i-x_0)^{\alpha}(y_i-y_0)^{\beta }\\
\end{array}
\end{equation}
of measured $ \bar {x} _i $, $ \bar {y} _i $ coordinates at Ep2 over
cross-moments of coordinate differences $x_i-x_0 $, $y_i-y_0 $ at Ep1
of a standard frame,
and where the index $i=0 $ again refers to the target.
Equation for $y $ has a similar structure.
A residual of  conditional equations at $i=0$ is treated as a proper motion
\begin{equation}
\label{eq:m2}
\mu_x= \bar{x}_0(Ep2)-x_0(Ep1)- A_0
\end{equation}
Note that all coefficients $A_m ^ {\alpha, \beta} $ with $m > 0 $  had vanished
since Eqs.(\ref {eq:r}) are written for a chosen $i=0$ star.
This result is completely equivalent to Eq.(\ref {eq:m}) since, 
multiplying Eq.(\ref {eq:r}) by $a_i $ and than summing up over $i $, 
in view of conditions (\ref{eq:t3}) we  find that
$A_0=N ^ {-1} \sum a_i (\bar {x} _i - x_i) $. Thus, 
symmetrization allows both to reduce the image motion and  exclude
all geometric distortions (their change between Ep1 and Ep2) 
up to the polynomial degree of
$ \alpha +\beta = k/2 -1 $ inclusive. 
And on the contrary, if the reduction model
(\ref {eq:r}) includes all  
modes $m \leq k/2 -1 $  with no omissions,
the least squares solution of 
systems (\ref {eq:r}) and (\ref {eq:t2} -\ref {eq:t3}) are {\it equivalent}. 
It means that though the use of the model (\ref {eq:r}) is aimed at
correction of geometric distortions, it reduces the image motion equally well.
For example, a 6 parameter linear model normally used in proper
motion works (e.g. Pravdo \& Shaklan \cite{Pravdo}) in fact performs 
the $k=4$ order symmetrization.

\section{Differential positions} 
A one-hour serie of $j=1,2 \ldots 72$ frames
obtained near the meridian was used for a detailed analysis.
A frame $j=1$ obtained near meridian 
was chosen as standard "epoch 1" frame. In this frame,
asuming some star as a target $i=0$ and using Eqs.(\ref {eq:t3}),
we computed a set of weights $a_i $ with which the target star positions
$W_j$ at each other $j$ frame ("epoch 2") were calculated.
Computations were performed at $k=2,4.. 12 $ and reference
field radii $R $ from 0.2$\arcmin $  to 1.1$\arcmin $, whereupon
all procedure was repeated,  considering in turn each other star as a target.
Since the actual image motion for so short time is zero,
the displacements computed, of course, 
represent errors of  $W_x$, $W_y$ 
determination. Because  a noticeable
part of  these errors are caused by centroiding errors  of reference
stars, we performed iterative refinement of  positions $ \bar {x} _i $,
$ \bar {y} _i $.
For this purpose we  carried out calculation of preliminary
$ W_x $, $ W_y $ values 
with $k=6 $ and near optimal $R=0.6\arcmin $. 
These displacements taken with reference to their average 
$\langle W_x \rangle $, $\langle W_y \rangle $
over the whole serie 
of $j=1,2 \ldots 72$ frames 
were treated as centroiding errors and therefore 
subtracted from  $ \bar {x} _i $,
$ \bar {y} _i $ coordinates of the corresponding $i$-th star. 
While reducing random errors of reference star positions,
these iterations introduce an extra atmospheric dependent 
and spatially-correlated bias. For this reason 
only a small improvement of the final precision was obtained  and 
only  at the first iteration; 
no improvement  was found for $k=2 $ at all.

\subsection {Astrometric error components}
\begin{figure}[htb]
\begin{tabular}{@{}c@{}c@{}}
{\includegraphics*[bb = 52 60 220 171, width=4.3cm,height=3.5cm]{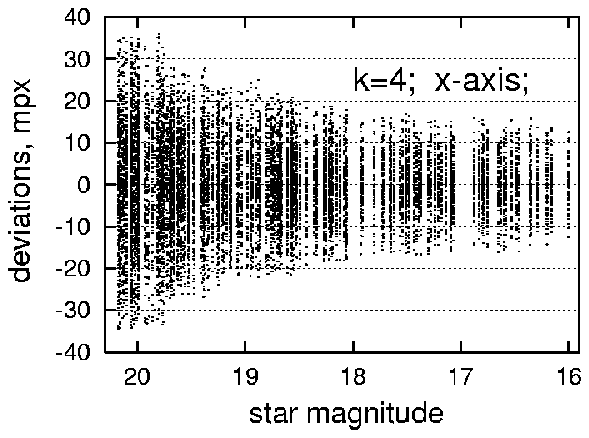}}&
{\includegraphics*[bb = 52 60 220 171, width=4.0cm,height=3.5cm]{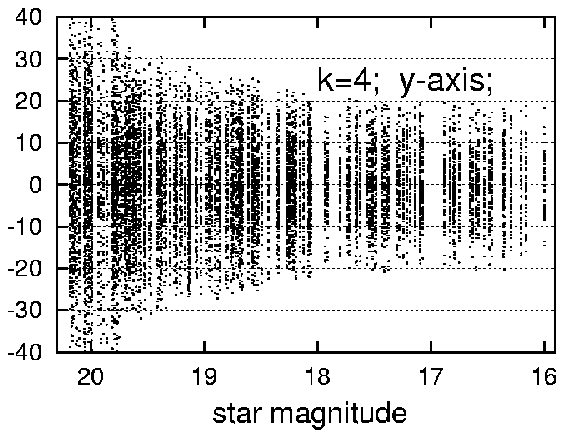}} \\
{\includegraphics*[bb = 52 48 220 171, width=4.3cm,height=3.9cm]{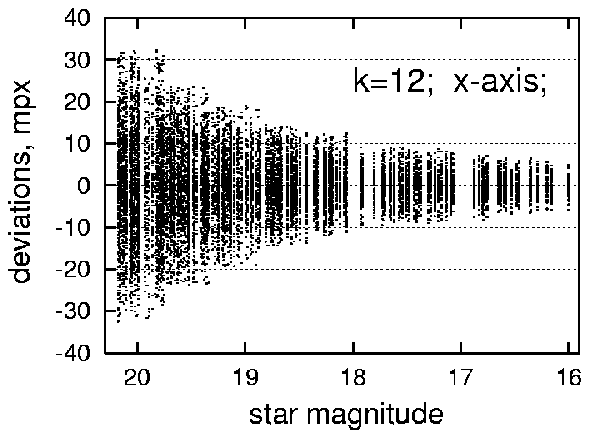}}& 
{\includegraphics*[bb = 52 48 220 171, width=4.0cm,height=3.9cm]{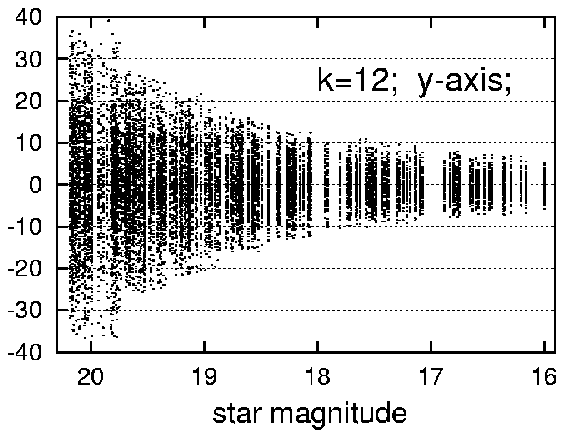}} \\
\end{tabular}
\caption {Astrometric deviations  $W_j -\langle W \rangle $ of each frame  star
position  $W_j$ from this star average position $\langle W \rangle $  as a 
function of instrumental 
magnitude; plots correspond to $0.8'$ reference field radius  and
reduction parameters $k=4$ and $k=12$
}
\label{fnew}
\end{figure}


Fig.\ref{fnew} represents astrometric errors as residuals 
$W_j -\langle W \rangle $
of each frame star position $W_j$ from its average $\langle W \rangle $
over all 72 frames. 
Distribution of residuals computed with $R=0.8'$, $k=4$ and $k=12$
is seen to be symmetric and dependent
on the star brightness which is especially contrast at $k=12$.
Errors of $W_y$ are slightly larger due to the image elongation along the 
$y$-axis. Note  the difference of point scatter at $k=4$ and $k=12$ which
were obtained with use of identical sets of reference stars.

The measured variance  $\Delta_{\rm{k}}^2 $ 
of each star positions computed with application of some $k$
order was found as  
$\Delta_{\rm{k}}^2 =\overline {(W_j -\langle W \rangle )^2}$.
It should be
commented that $\Delta_{\rm{k}}$  defines also a precision of $W$ value
zero-points which are set so that $W(Ep1)=0$. The $W(Ep1)$ 
values are thus shifted
from $\langle W \rangle$ randomly by about $\Delta_{\rm{k}} $  r.m.s.

\begin{figure}[htb]
\begin{tabular}{@{}c@{}}
{\includegraphics*[bb = 52 69 294 221, width=8.0cm, height=5.0cm]{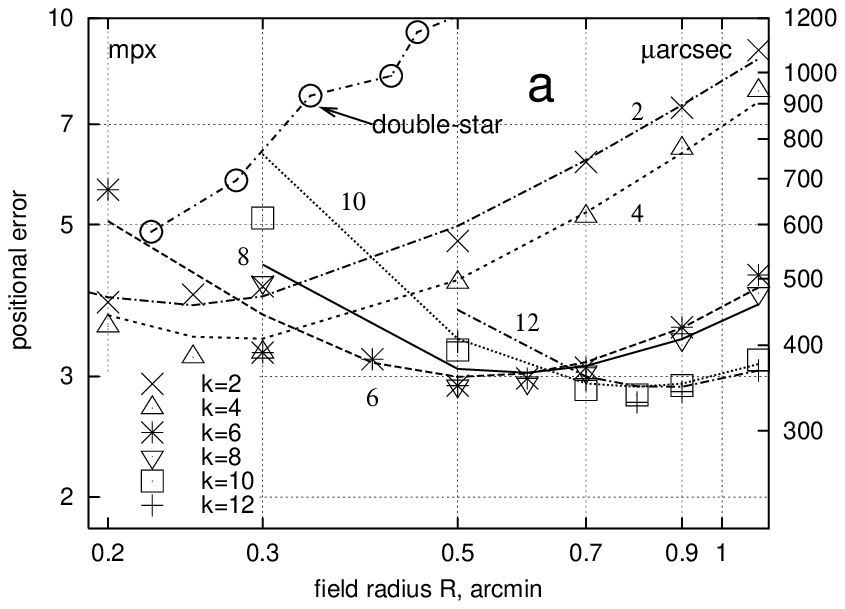}} \\
{\includegraphics*[bb = 52 69 294 221, width=8.2cm,height=5.0cm]{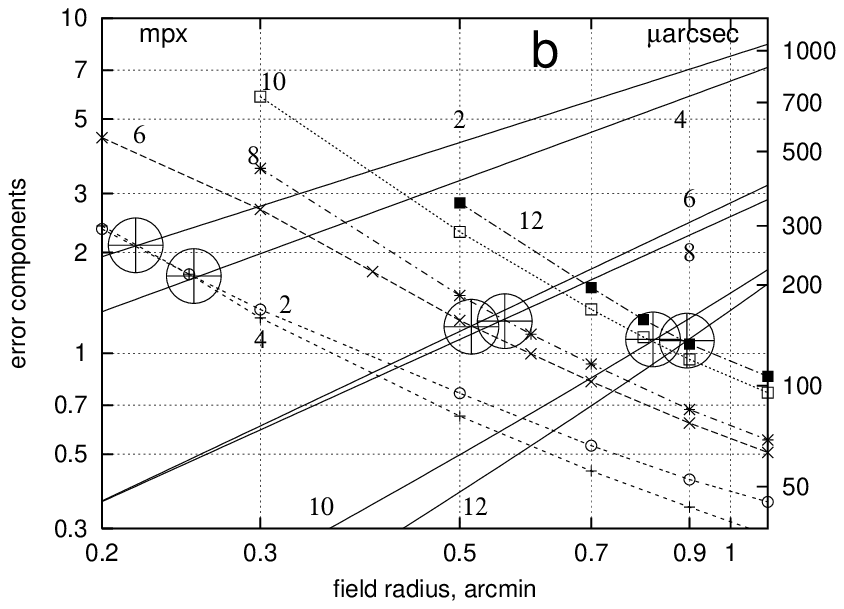}} \\
{\includegraphics*[bb = 52 48 294 221, width=8.0cm,height=5.5cm]{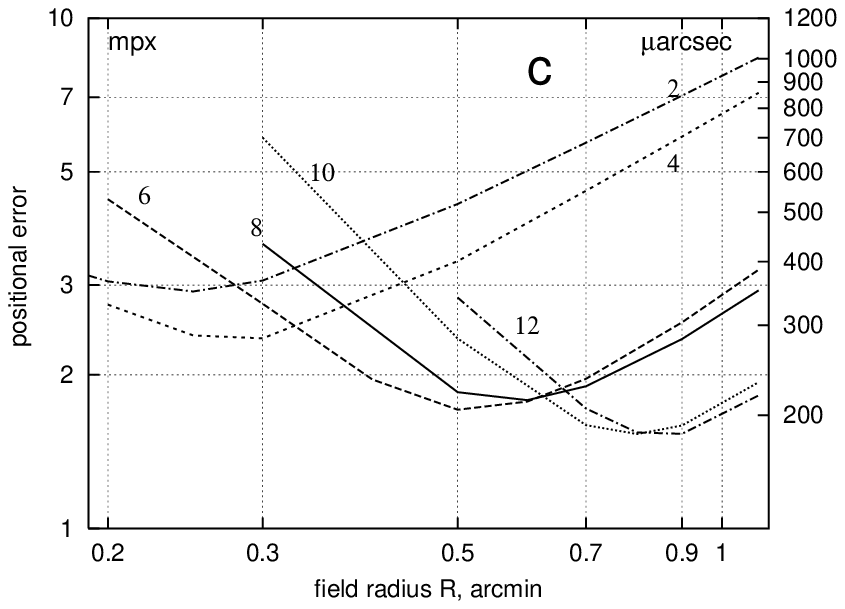}}\\ 
\end{tabular}
\caption { Dependence of the total astrometric error and its components on $R$ 
at $k=2,4..12$;
{\bf a)} the total error $\Delta^2_{\rm{k}}$ of 16.5 mag  star observation:
measured point estimates (dots) and  approximation (\ref{eq:decomp}) (lines); 
errors of a double-star separation measurements are shown with a shift:
a real curve is a factor 2 upwards;
{\bf b)} reference field components:  $\sigma_{\rm{rf}}$
(dashed lines with different symbols) and  $\sigma_{\rm{at}}$ (solid lines);
Large cicles mark points of line intersections at optimal radii   
$R_{\rm{opt}}$;   
{\bf c)} the cumulative reference field contribution
$ \Delta _ {\rm {a+r}} $. All plots refer to right ascention}
\label{f1}
\end{figure}

The estimates of $\Delta^2_{\rm{k}}$ plotted as a function of $R$
show a large scatter of points (Fig.\ref{f1}a) however a clear dependence 
on $k$, $R$
and, of course, on the star brightness (the plot is built for 16.5 mag
star) is seen. 
To model this dependence, we decomposed the $\Delta^2_{\rm{k}}$ value
as a sum
\begin{equation}
\label{eq:decomp}
\Delta^2_{\rm{k}}= \sigma^2_{\rm{at}}+ C^2\sigma^2_{\rm{rf}} + \eta^2
\varepsilon_0^2 
\end{equation}
of atmospheric component (\ref {eq:t1}) which at large number of reference
stars is approximated by the expression (Eqs.(36--38) of Paper I) 
\begin{equation}
\label{eq:atm}
\sigma_{\rm{at}}=B_k (R/1\arcmin)^{b_k}
\end{equation}
with  parameters $B_k $ and $b_k $  dependent on $k $ and considered here as
free model parameters;
centroiding noise of reference field
\begin{equation}
\label{eq:srf}
\sigma_{\rm{rf}}=N^{-1}\sqrt{\sum a_i^2 \varepsilon_i^2}
\end{equation}
where $ \varepsilon_i$ is given by (\ref {eq:accur});
and the third component $\eta \varepsilon_0$
which represents the centroid precision for the target star.
The coefficient $C < 1 $ is introduced to reflect the decrease of 
$\sigma _ {\rm {rf}} $ component produced by  iterative refinement of 
star positions, 
and $ \eta \geq 1 $  is the ratio of the true centroid error 
to its  photon-limited estimate (\ref{eq:accur}); it was treated as a free
model parameter for each 0.5 mag bin of magnitudes.
Solution of Eq.(\ref {eq:decomp}) produced 
$B_k $ and $b_k $ values,  $C\approx 0.7 $,  $ \eta \approx 1 $
for stars fainter than 17~mag and, due to PSF profile imperfection,
$ \eta ^2 \approx 1.2-1.3$  for brighter images.

Fig.\ref{f1}b shows dependence of  components
$C\sigma _ {\rm{rf}} $ (units: 0.001~px = 1~millipixel, hereafter mpx; 
1~mpx = 125~$\mu$as (microarcsec)) and $ \sigma _ {\rm {at}} $ on $R $ at 
each  parameter $k $ value.
The observed dependencies qualitatively match theoretic behavior of
plots shown 
in Fig.15 of Paper I and correspondent to $D=10$~m and $T=10$~min.
At narrow fields, the dominant component is $ \sigma _ {\rm {rf}} $ which 
decreases with $R $  following an approximate dependence (Paper I)
\begin{equation}
\label{eq:a}
\sigma_{\rm{rf}}= \frac{\mbox{FWHM}}{2.36R\sqrt{\pi n}} \frac{k}{4},
\end{equation}
where $n $ is the total light from a  1x1 $\arcmin $ reference field,
$R $ is given in minutes of an arc,
$k/2 $ is supposed even and star number $N \gg N _ {\rm {min}} $.
For small $N \simeq 3-5 N _ {\rm {min}} $, the actual values (\ref{eq:srf})
of $ \sigma _ {\rm{rf}} $ are systematically a factor 1.5--2 above
the estimate (\ref{eq:a}). At yet lower $N$ the curves for $\sigma _ {\rm {at}}$
are not built.
At $R \simeq 1 \arcmin $ the value of $ \sigma _ {\rm {at}} $ considerably
exceeds $ \sigma _ {\rm {rf}} $.

A twinning  of curves in  Fig.\ref{f1} 
related to pairs $k $ = (2,4), (6,8) and (10,12)  is a 
specific feature  of densely populated fields since
a  high sky star density  favors reduction due to the field averaging effects.
At $k$ not multiple 4 the values of $ \sigma _ {\rm {at}} $ and
$ \sigma _ {\rm {rf}} $ are therefore close to  approximate 
estimates (\ref {eq:atm}) and (\ref{eq:a}) computed with  $k+2 $.
Just for illustration of the field  averaging effect,  Fig.\ref{f1}a reproduces
the dependency of astrometric error for measurements of 
$\bar {x}_0-\bar {x}_i$ separation  between two stars of approximately equal 
brightness.
The dependance observed implies that to obtain a good astrometric precision,
a  separation  $R$ between two stars should be less than 0.1$\arcmin$; 
at larger $R$ the accuracy degrades very rapidly.
This is an example of a marginal single-star reference field but note that
$W_x$ positions (\ref {eq:t2}) are composed of  $\bar {x}_0-\bar {x}_i$
unit differences averaged over the field.

\begin{table}[tbh]
\caption [] {Parameters $B_k $ [$ \mu \mbox {as} $] and $b_k $
calculated for $X $ and $Y $ and scaled from Fig.15 of Paper I}
\begin{flushleft}
\begin{tabular}{r|rr|rr|rr}
\hline
\hline
& $B_k$ & $b_k$ &  $B_k$ & $b_k$ & $B_k$ & $b_k$\\
$k$ & \multicolumn{2}{c|}{ computed in X}  
&\multicolumn{2}{c|}{ computed in Y } 
&  \multicolumn{2}{c}{ from Paper I}  \rule{0pt}{11pt}\\ 
\hline
2 &  1380  &  0.9 &  1260  &  0.8    &  1190  &  0.64  \\
4 &  780  &  1.0 &  1010  &  1.0    &   806  &  1.69 \\
6 &  340  &  1.3  &  420  &  1.2    &    186  &  1.63  \\
8 &  310  &  1.2  &  370  &  1.1    &    161  &  1.83  \\
10 &  180  &  1.8  &  230  &  1.4    &    84  &  1.66   \\
12 &  160  &  1.9  &  200  &  1.6    &    76  &  1.84   \\ 
\hline
\end{tabular}
\end{flushleft}
\end{table}

Table 1 compares current parameter $B_k $ and $b_k $ estimates for VLT
and their expectations
for Chilean observatories (Cerro Tololo, Paranal and San Pedro Martir) 
at  typical  atmospheric conditions.
This data was found by scaling plots from Fig.15 of Paper I given for
a $D=10 $~m and $T=10$~min, to the current values using 
a relation 
$B_k^2 \sim {\mbox {(airmass)}} \times T^{-1}D^{-3}$ (  Eq. (37) of Paper I).
Results given in the last column of Table 1 generally agree with observed
values except some differences commented below.
By definition (\ref {eq:atm}),  coefficient $B_k $ is equal to 
$ \sigma_{\rm {at}} $ at $R =1\arcmin $.
Its measured value is approximately equal to the scaled expectation 
at $k=2 $ and $k=4$ but starting from $k=6 $ is a factor 2 -- 3 higher.
This may indicate an actual intensity of  atmospheric turbulence
at high altitudes
which, according to the discussion in Paper I, may easily show such 
fluctuations.  
A smaller than expected measured value of coefficient $b_k $ 
at $k=4 $, 6 and 8 implies a higher value of 
$ \sigma _ {\rm {at}} $  value at small $R $; 
unlikely it is of atmospheric origin and probably
is due to some effects caused by the active optics performance.

Let us consider
a sum $ \Delta^2 _ {\rm {a+r}} = \sigma _ {\rm {at}} ^2 +
C^2\sigma^2 _ {\rm {rf}} $ that determines the properties of a noise from
reference field only.
This component  (Fig.\ref{f1}c) has a typical 
parabola-like shape with a minimum  at the point $R _ {\rm {opt}}$ 
where  $\sigma _ {\rm {at}} ^2=C^2\sigma^2 _ {\rm {rf}} $.
Note a gradual decrease of $ \Delta _ {\rm {a+r}} $ value 
with applying large $k $ and a corresponding increase of $R _ {\rm {opt}}$. 
In our case, the minimum is achieved at $k=10, 12 $ and $R \sim 0.9 \arcmin $,
where
$ \Delta _ {\rm {a+r}} = 1.59 $ mpx for right ascension and $ \Delta _ {\rm
{a+r}} = 1.91 $ mpx for declination. The total astrometric error
$ \Delta _ {\rm {k}}$ changes with  $R$ similarly and also has a minimum
at   $R=R_{\rm {opt}}$. 
                                                        
Centroid errors $\eta \varepsilon$ for the brightest unsaturated stars 
of 16~mag are equal to 
1.86 mpx (230 $\mu$arcsec)  for right ascension and 
2.22 mpx (280 $\mu$arcsec)  for declination 
and are of the order of best attainable $ \Delta _ {\rm {a+r}} $.
The atmospheric error components, $ \Delta _ {\rm {k}} $ value 
for these stars and $R _ {\rm {opt}} $ for each $k $ 
are listed in Table 2.


\begin{table}[tbh]
\caption [] { Astrometric error components (mpx) and
optimal  field radii $R _ {\rm {opt}} $  at $k=2,4 \ldots $; 
last column contains data averaged over $k$; centroid error 
$ \eta \varepsilon_0 $ is given for brightest stars}
\begin{flushleft}
\begin{tabular}{@{}r|rrrrrr|c@{}}
\hline
\hline
& \multicolumn{6}{|c|}{$k$ } &  \\ & 2 & 4& 6& 8& 10 & 12 & $\langle \rm{k}
\rangle $ \\ 
\hline
& \multicolumn{6}{|c|}{$x$, right ascension} &  \\
$\sigma_{\rm{at}}$ & 2.1&   1.7&     1.3    & 1.3   &  1.1  &  1.1    &  -   
\\ $C\sigma_{\rm{rf}}$ & 2.1&   1.7&     1.3    & 1.3   &  1.1  &  1.1    &  - 
  \\ $ \eta \varepsilon_0$ &1.9 & 1.9 &    1.9    & 1.9   &  1.9  & 1.9    & 
1.9   \\ 
\hline
$\Delta_{\rm{a+r}}$ & 3.0 &  2.4&    1.8 &    1.9   & 1.6   &  1.6   &
\multicolumn{1}{c}{$ \Delta_{\langle\rm{a+r}\rangle} =1.1 $}\\
$\Delta_{\rm{k}}$ & 3.5 &  3.0&    2.5 &    2.5   & 2.4   &  2.4   &
\multicolumn{1}{c}{$\Delta_{\langle \rm{k}\rangle} =2.2 $}\\ \hline
\hline
& \multicolumn{6}{|c|}{$y$, declination }& \\
$\sigma_{\rm{at}}$ & 2.8&   2.0&     1.5    & 1.6   &  1.4  &  1.4    &   -  
\\ $C\sigma_{\rm{rf}}$ & 2.8&   2.0&     1.5    & 1.6   &  1.4  &  1.4    &  - 
  \\ $ \eta \varepsilon_0$ &2.2& 2.2 &    2.2    & 2.2   &  2.2  & 2.2    & 
2.2   \\ 
\hline
$\Delta_{\rm{a+r}}$ & 4.0 &  2.8&    2.1 &    2.2   & 2.0   &  1.9   &
\multicolumn{1}{c}{$ \Delta_{\langle\rm{a+r} \rangle}=1.4 $}\\
$\Delta_{\rm{k}}$ & 4.6 &  3.6&    3.0 &    3.1   & 2.9   &  2.9   &
\multicolumn{1}{c}{$ \Delta_{\langle\rm{k}\rangle} =2.6 $}\\ 
\hline
\hline
$R_{\rm{opt}}$ & 0.22$\arcmin$ &  0.28$\arcmin$  &  0.5$\arcmin$ &
0.55$\arcmin$ &0.8$\arcmin$ &0.9$\arcmin$&  -    \\ \hline
\end{tabular}
\end{flushleft}
\end{table}

Note that for any arbitrary $k $ value  (or equivalent order
$m=k/2-1 $ of a polynomial model (\ref {eq:r})), one can find
always  a field size $R=R _ {\rm {opt}} $ at which $ \sigma _ {\rm {at}} =
C\sigma _ {\rm {rf}} $ and where atmospheric image motion 
represents less than a half of the 
total variance (\ref{eq:decomp}). Astrometric measurements 
thus are {\it never} atmospherically limited providing that 
the reduction is performed with  $R_{\rm {opt}} $. 
This conclusion is valid also for the sky areas with a poor
sky star density as discussed in Sect.5.                                

\subsection {Averaging over modes} 
Positions $W_k $ calculated with different
$k $ and different reference field radii $R _ {\rm {opt}} $
are related to distinctive sets of stars and hence have almost 
uncorrelated components $ \sigma _ {\rm {rf}} $. Atmospheric errors 
 $ \sigma _ {\rm {at}} $ computed with different $k $  
(modes $m $ in Eq. (\ref {eq:t1})) 
are also weakly correlated since the effective turbulent layer heights $h$
depend on $k$. Thus, while at $k=2$ all turbulent layers from 5 to 20~km heights 
contributes equally to the variance of $W_k $, the use of 
$k=12 $ effectively cut off turbulence contribution from $h<20$~km 
(Fig.12 of Paper I). It is natural therefore to average positions $W_k $ 
with weights $P_k=1/\Delta _ {\rm {a+r}} ^2 $ calculated  based on Table 1: 
\begin {equation}
\label{eq:w}
W_{\langle k \rangle}=\sum_k P_{k}W_{k} /\sum_k P_{k}
\end{equation}
The total weight of the average is $ \sum P_k=4.0 P _ {12} $.
However, taking into account almost equal size of
reference fields for pairs $k $ = (2,4), (6,8) and (10,12), the effective weight 
is rather $P _ {\rm{eff}} \simeq 2.0 P _ {12} $. The variance of 
weighed positions (\ref {eq:w}) is
\begin{equation}
\label{eq:finerr}
\Delta^2_{\langle k \rangle}= \Delta^2_{\langle \rm{a+r} \rangle}
+ \eta^2 \varepsilon_0^2
\end{equation}
with a new  reference field component
\begin{equation}
\label{eq:rf}
\Delta^2_{\langle {\rm{a+r}} \rangle}=\frac{ P_{12}}{P_{\rm{eff}}}
\Delta^2_{ \rm{a+r}} \left |_{\rm{\;at  \; k=12}} \right. 
\end{equation}
decreased almost twice (Table 2) in comparison to its value
$ \Delta^2 _ {\rm {a+r}} $ at $k=12 $.

\begin{figure}[htb]
\centerline{\includegraphics*[bb = 52 58 294 222, width=8.0cm]{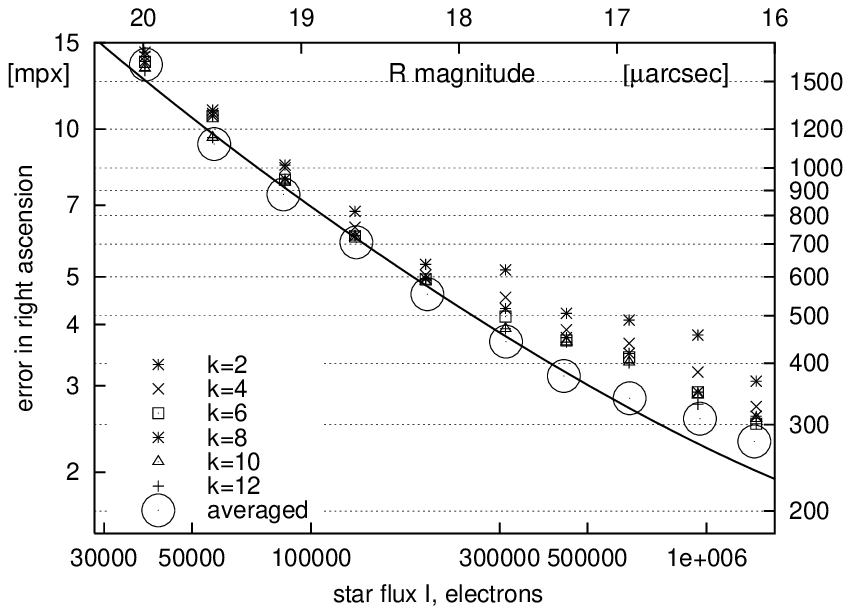}}
\centerline{\includegraphics*[bb = 52 48 294 222, width=8.0cm] {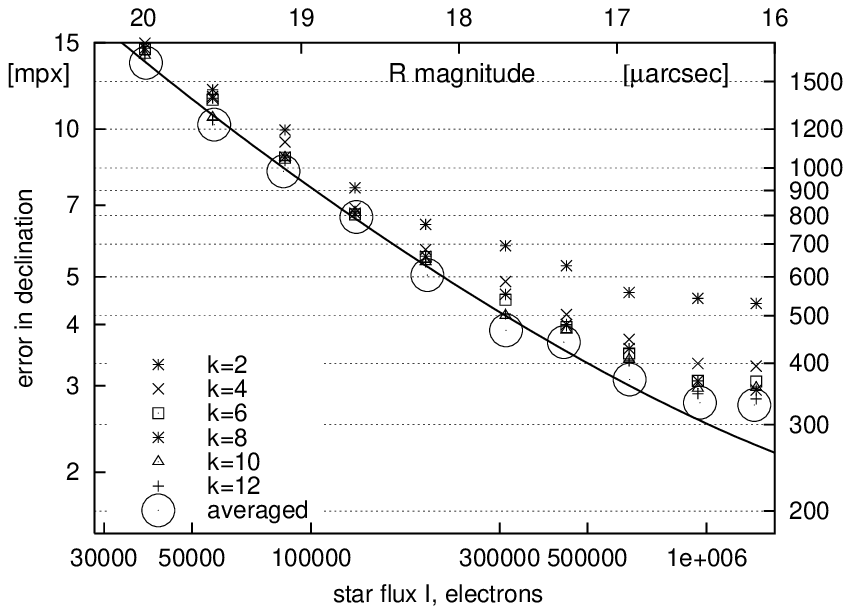}} 
\caption{Astrometric error as a 
function of the target flux $I $. Different dots -- 
 $ \Delta _ {\rm {k}} $ at 
$k=2,4 \ldots 12 $; open circles -- error $ \Delta _ {\langle k \rangle} $ 
of positions averaged over modes; solid line -- approximation
(\ref {eq:finerr}) of $ \Delta _ {\langle k \rangle} $ asuming
$ \eta=1 $ and with $ \varepsilon_0 $ given by Eq.(\ref {eq:accur}).
{\it Upper panel} -- right ascention, { \it lower}  -- declination}
\label{vxy}
\end{figure}


Fig. \ref {vxy} shows errors $ \Delta _ {\rm {k}} $ (at $R=R _ {\rm {opt}}
$) and $ \Delta _ {\langle k \rangle} $ as a function of 
the target brightness. These estimates were obtained from the
convergence of the
measured positions $W_k $ and $W _ {\langle k \rangle} $ at 
different frames and after applying a small correction for color
effects (Sect.6).
The model dependence (\ref {eq:finerr}) plotted at $ \eta=1 $ (solid
curve) predicts well the measured estimates till
$R<$17~mag. At brighter end the actual errors exceed expected at $ \eta=1 $
due to non-Poisson noise in  star profiles ($ \eta ^2\sim 1.2 $).
Roughly, the astrometric error $ \Delta _ {\langle k \rangle} $ is equal
to $ \varepsilon_0 $ at $R > 17 $~mag and to 
$ \ 1.5 \varepsilon_0 $ for brighter
stars. An error component introduced by incorrect 
tying to the reference frame thus is rather small.

\section {Astrometric error at low sky star density}

The results obtained above  for high sky star density frames can be
used  to predict astrometric performance of VLT at other sky regions.
The most important parameter necessary for this purpose is the 
integral light flux $n$ registered  from reference stars.
Its average value for our OGLE field is equivalent to $n_0=12\times 10^6 $~e$ ^- $ 
from a square minute of arc (60 stars brighter than $R=20$~mag at 
$1 \times 1\arcmin $ area).
It essentially affects $ \sigma _ {\rm {rf}}$ value which  according
to Eq.(\ref {eq:a})  goes as $ \sigma _ {\rm {rf}} \sim n^{-1/2} $ 
while $ \sigma _ {\rm {at}} $  does not
depend on $n $ (\ref {eq:atm}). 
At $n \neq n_0 $ the curves in Fig.1 relevant to the  
$ \sigma _ {\rm{rf}} $ component  are shifted moving the point 
$R _ {\rm {opt}} $ to a new
position $R ' _ {\rm {opt}} = \theta R _ {\rm {opt}} $.


\begin{figure}[htb]
\resizebox{\hsize}{!}{\includegraphics{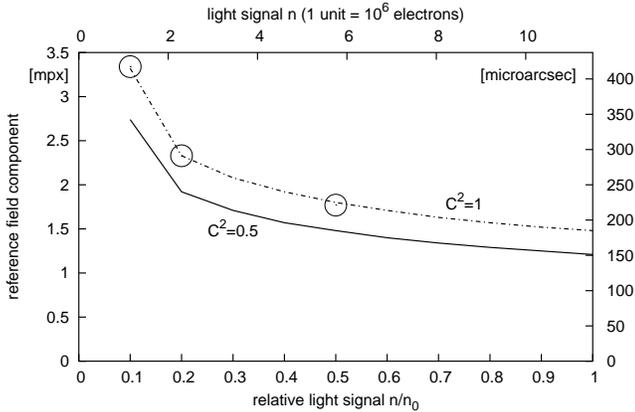}}
\caption {An error of referencing  $ \Delta _ {\langle \rm {a+r}\rangle} $ 
as a function of reference field total light $n$:
based on the scaling expression (\ref {eq:u2}) at $C^2=0.5 $ (solid curve),
the same at $C^2=1 $ (dashed line) and  found from a direct
reduction of a sparsed original sample of stars (open circles)}
\label{vn}
\end{figure}

Using Eqs.(\ref {eq:atm}) and (\ref {eq:a}) for scaling  
$ \sigma _{\rm {at}} $ and $ \sigma _ {\rm {rf}} $ components at
a new value of  $R ' _ {\rm {opt}} $, from a requirement of a balance 
$\sigma _ {\rm {at}} = C \sigma _ {\rm {rf}} $ we find  
$ \theta = [(\omega C^2 n_0) / (C^2_0 n)] ^ {1 / (2b_k+2)} $ where $C_0^2=0.5 $.
Note that Eq. (\ref {eq:a}) gives systematically underestimated 
$\sigma _ {\rm {rf}} $ value for frames with a small number of reference stars 
$N \sim(3-5) N _ {\rm {min}} $. 
In view of this we introduce the parameter $ \omega \geq 1 $ equal to the
ratio of the actual $ \sigma _ {\rm {rf}} $ value 
given by Eq. (\ref {eq:srf}) to its approximate estimate 
(\ref{eq:a}). For the sample  analyzed, $ \omega \simeq 1 $ at 
$n \geq 0.2 n_0 $ and at low
$n=0.1 n_0 $ it grows from 1 to 2.5 when $k $ changes from 4 to 12.
For $R _ {\rm {opt}} $ and $b_k$ relevant to some $k $ we obtain 
$ R'_{\rm{opt}} =  [(\omega C^2 n_0)/(C^2_0 n)]^{1/(2b_k+2)} R_{\rm{opt}} $
hence the expected error of  referencing
\begin{equation}
\label{eq:u2}
\Delta'_{ \rm{a+r} } =  [(\omega C^2 n_0)/(C^2_0 n)]^{b_k/(2b_k+2)} 
        \Delta_{\rm{a+r} } 
\end{equation}
Based on this expression and using parameter values listed in Tables 1 and 2, 
we have calculated the expected values of $ \Delta ' _ {\rm {a+r}} $ at 
$n=0.1 \ldots 1.0n_0 $ in two variants, with $C^2=0.5 $ and with $C^2=1 $.
The resulting error $\Delta _ {\langle k \rangle} $ 
that corresponds to the error (\ref {eq:rf})  of the average over modes 
is shown in Fig.\ref{vn}.

To control our scaling approach, we performed direct computations of 
$ \Delta _ {\langle k \rangle} $ with use of the original  but 
artificially sparsed star sample and with no iterative improvement of reference 
star positions ($C^2=1 $).
Results (open circles) match well the estimates (\ref{eq:u2}).
Note that  at $n \geq 0.2n_0 $ a good referencing at the  1.5 - 2~mpx level 
is maintained, which is comparable 
to the bright star centroid precision. At $n \sim 0.1n_0 $ due to the
decrease of a light signal from reference stars, the accuracy of the 
referencing sharply degrades. 
Besides, at  highest $k=12 $ order the optimum  field size 
$2R'_ {\rm {opt}} = 3.2-3.5\arcmin $ is near the  CCD size.

At very low $n$ the star number in the area of the optimal radius
$R=R'_ {\rm {opt}} $ may be smaller than $N_ {\rm {min}} $ and so
be insufficient for implementation of the desired $k$ order. This is
extremaly unfavorable since forces one to use lower $k$ or, even worse,
performe reduction with not optimal $R$. Using above expressions for 
$R'_ {\rm {opt}} $ and  $N_ {\rm {min}} $ and asuming $C=C_0$ we find that
a low limit of reference star density necessary to implement some $k$ 
symmetry order is
$n_{\rm{min}}=n_0 [k(k+2)/(8\pi n_0 R^2_ {\rm {opt}})] ^{1+1/b_k} \omega ^{-1/b_k}$
Asuming $\omega \approx 2-3$ and taking $b_k$ and  $R_ {\rm {opt}} $ values 
from Tables 1 and 2, we find that approximately
$n_{\rm{min}} \simeq 0.01 n_0$ at any
$k$ from 4 to 12. This integral star density (about 0.6 stars brighter 
than R=20~mag per $1 \times 1\arcmin $ area) 
is typical to the Galactic pole (Allen \cite{Allen}).

Note that  the unequality  $N>N_ {\rm {min}} $ can be easily restored
in any case by incorporation of some extra faint stars as reference. 
This does not degrade precision, on the contrary,
the total flux $n$ will increase only and, 
according to Eq.(\ref {eq:a}),
$\sigma _ {\rm {rf}}$ improves. We  conclude therefore that even
at low sky star density the key problem of astrometric error improvement
is related to the telescope light signal limitation but the atmospheric image
motion (see Table 3, last column).

\begin{table}[tbh]
\caption [] {Centroid precision and astrometric error ($ \mu$as) at 10~min 
total exposure and different $n/n_0$ for 
reduction versions with $C^2=1.0 $ and 0.5.
A ratio of $\sigma^2_{\rm{at}}$ to the total error 
(last column) and  reference field 
contribution $\Delta'_{ \rm{a+r} }$ (last line) are given also}
{ \small
\begin{flushleft}
\begin{tabular}{@{}r|r|rrr|rrr|r}
\hline
\hline
& centr& \multicolumn{3}{|c|}{red. with $C^2=1$}& \multicolumn{3}{|c|}{red.
with $C^2=0.5$}&   $\sigma^2_{\rm{at}}$\\ 
$R$ & error&  \multicolumn{3}{|c|}{$n/n_0$} &
\multicolumn{3}{|c|}{$n/n_0$} & ratio \\ 
mag  & $\eta \varepsilon$ &
             0.1 &  0.2 & 1.0 & 0.1 & 0.2 & 1.0 & \% \rule{0pt}{11pt}\\ 
\hline
20 &  300 &  308 &  304 &   301 &   305 &  302&  301 & 0\\
19 &  167 &  181 &  174 &   170 &   177 &  172&  169 & 1\\
18 &   98 &  120 &  110 &   103 &   114 &  106&  101 & 3\\
17 &   65 &   95 &   81 &    72 &    87 &   77&   70 & 6\\
16 &   42 &   81 &   64 &    52 &    71 &   58&   49 & 13\\
15 &   26 &   75 &   56 &    41 &    63 &   48&   36 & 24\\
14 &   16 &   72 &   52 &    35 &    60 &   44&   30 & 35\\ 
\hline
(0) &(0) &    70 &   49 &    31 &    58 &   40&   25 & 50\\ 
\hline
\end{tabular}
\end{flushleft}
}
\end{table}

Table 3 gives the centroid precision and astrometric error 
(\ref{eq:finerr}) for 14 -- 20 mag (R)  stars expected from a reduction of 
a serie of frames with  a total 
$T=10 $~min exposure and of  30 -- 50~min (depending on observation mode) 
duration. For bright 14 -- 16 mag  stars we adopted $ \eta^2=1.3 $, 
for 17~mag we used $ \eta^2=1.2 $, for fainter stars $ \eta=1 $. The
estimates are given for reduction versions correspondent to $C^2=1 $ and
$C^2=0.5 $. In the last line, the total reference field contribution  
$\Delta'_{ \rm{a+r} }$ which is equal to  the  limiting
astrometric precision of very bright stars ($ \varepsilon =0 $) is given.
At high sky star density, this limit is 25~$\mu$as per a 10 min exposure. 
The last column contains a ratio of atmospheric image motion to the
total error; as discussed above, this portion cannot exceed 50\% by
the definition of the optimal field size.

For brightest stars, thus, the expected accuracy of observations 
in Galactic bulge is equal to 30--50~$\mu$as at 10~min
exposure, degrading to 50--80~$\mu$as at $n \sim 0.1-0.2 n_0 $
(2--4 $\cdot 10^4 $ stars to $R=20$~mag per square degree which
corresponds to galactic latitudes of 20--30$\degr $ (Allen \cite{Allen})). 
For stars fainter than $R > $17~mag, the error does not depend on
characteristics of reference field due to a small light signal from the target.

\section{DCR and residual chromatism of LADC } 
The effects of DCR was considered elsewhere. In particular,
Monet et al. (\cite {Monet}) studied 
restriction of the accuracy of proper motions derived
from the reduction  of frames obtained at 
different hour angles. In this study we use mathematical approach similar 
to that used by Monet et al. (\cite {Monet}).
However, we have to take 
account of joint influence of an atmosphere and of chromatic corrector 
LADC (Avila et al. \cite {Avila})  used at VLT to improve the quality
of images.

It is known that the relative displacement $S $ of two monochromatic images with
wavelengths $ \lambda_1 $ and $ \lambda_2 $ is proportional to a tangent
of zenith distance $z $: $S =\delta \tan z $ where $ \delta $ is the 
coefficient of atmospheric
dispersion dependent on a difference $ \lambda_1 - \lambda_2 $.
The corrector reduces this displacement, incorporating a negative 
equivalent displacement $ -(\delta -d) \tan z $ where $d $ is the
residual error of the corrector. The
observed relative displacement of the images in a vertical direction 
therefore is  $d\tan z $. 
When a serie of frames is obtained at VLT, the  corrector
is preset at some fixed zenith distance $z _ {\rm {L}} $ and this 
setting does not change. Therefore
\begin{equation}
\label{eq:ladc1}
S = \delta (\tan z - \tan z_{\rm{L}}) + d \tan z_{\rm{L}}
\end{equation}
If $ \lambda_1 $ is an effective wavelength of a target star and $ \lambda_2 $
is effective wavelength of reference field, the measured displacements 
$S_x, S_y $ are equal to the  $S $ projection at the
CCD coordinate system axis
($x $: oposite to right ascension, $y $: along declination)
\begin{equation}
\label{eq:ladc2}
\begin{array}{lr}
S_x = S \sin \gamma +S_{x0}, & S_y = S_0 - S\cos \gamma +S_{y0},
\end{array}
\end{equation}
where $ \gamma $ is the angle formed by a direction to zenith and the $y $ axis, 
$S_0 $ is the $S $ value in a meridian and $S _ {x0} $, $S _ {y0} $ are 
zero--points.
            
\begin{figure}[htb]
\resizebox{\hsize}{!}{\includegraphics*[bb = 50 57 301 226 width=8.0cm]{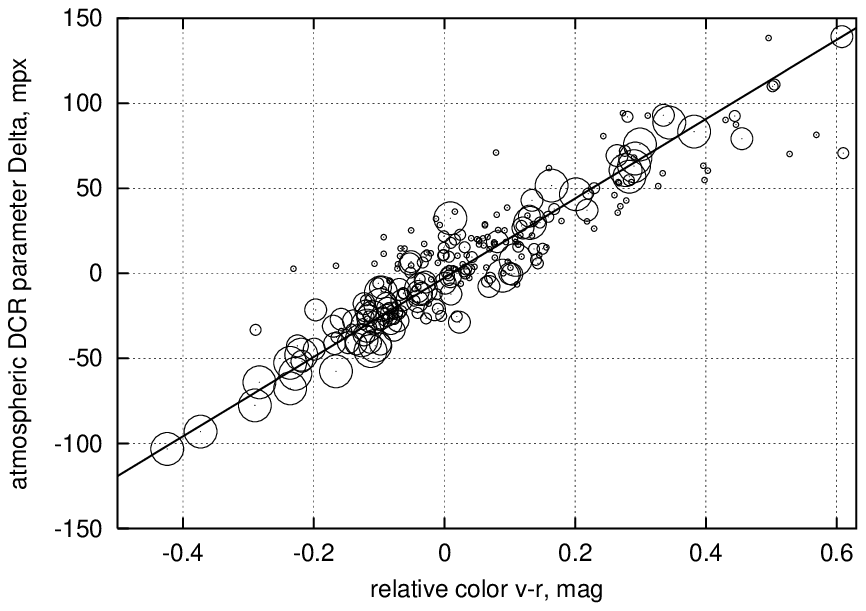}}
\resizebox{\hsize}{!}{\includegraphics*[width=8.0cm]{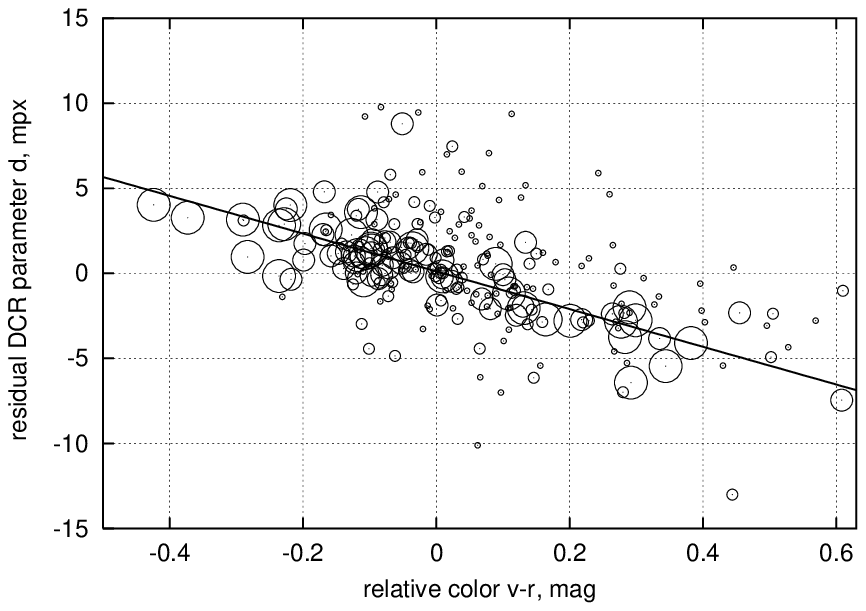}} 
\caption {Atmospheric dispersion coefficient $ \delta $ ({\it upper panel}) and
the residual LADC bias $d $ ({\it lower}) as a function of relative
star color $v-r-\overline {(v-r)} $; the size of 
circles is proportional to  the star brightness (from $3\cdot 10^4 $  to
$1.2\cdot 10^6$e $ ^- $); solid lines - approximation (\ref {eq:ladc3})
}
\label{coef}
\end{figure}

The values of  $ \delta $ and $d $ were found  from  a least squares 
solution of Eq.(\ref {eq:ladc2}) for each star.
As input, we used positions of stars in all 280 frames which were
calculated with $k=10 $ and $R=0.8\arcmin $.
The results are shown in Fig.\ref {coef} as a function of relative color
$v-r-\overline {(v-r)} $ where $ \overline {v-r} $ is the 
mean instrumental color index of a local star group used as reference 
for some target star. The distribution of point estimates in plots
is well approximated by linear functions
\begin {equation}
\label{eq:ladc3}
\begin{array}{l}
\delta = -2.6+(233 \pm 8)[v-r-\overline{(v-r)}] \\
d= -(11.0 \pm 0.9)[v-r-\overline{(v-r)}]
\end{array}
\end{equation}
where values are expressed in mpx and correspond to atmospheric
pressure 744~mb, temperature 11$\degr $ and 11\%  humidity.
Residual scatter of points in Fig.\ref {coef} is
$ \sigma (\delta) = 11.9 $~mpx for $ \delta $ and $ \sigma (d) =1.3 $~mpx
for $d $ and  a factor 4--5 exceeds  formal errors of 
$ \delta $ and $d $ value detrmination. Only  partially it can be
explained by the errors  of color determination
($ \pm 0.02 $~mag). Most likely, this scatter is intrinsic and
can not be adequately modelled using
only $v-r $ colors.

To check a reliability of the results obtained, we performed a control 
computation of a chromatic refractive index $\rho$ using 
the  expression 
\begin{equation}
\label{eq:dr}
\rho = 346 {\mbox {mas}} \int ^{\lambda_{max}}_{\lambda_{min}} \lambda ^{-2}
T(\lambda )I(\lambda) \, d \lambda / \int ^{\lambda_{max}}_{\lambda_{min}} T
(\lambda )I(\lambda) \, d \lambda
\end{equation}
given by Pravdo \& Shaklan (\cite {Pravdo}).
The integration is performed within the wavelengths  limits $ \lambda _ {min} $,
$ \lambda _ {max} $  defined by a combined transmission  $T (\lambda) $ of 
the filter and CCD and  $I (\lambda) $ is the star flux function of wavelength.
This expression is normalized
to 0$\degr $ and pressure of 1013~mb. We calculated 
$ \rho$ value for stars with blackbody temperatures corresponding to 
G0 and M0 spectral types which $V-R$ colors
0.52 and 1.1 are typical to the analyzed sample of stars. The difference of
$ \rho $ values computed for these spectra
and  reduced to a unit interval of colors is  38~mas/mag and
corresponds to the color coefficient  in expression (\ref {eq:ladc3}) for
$ \delta $. After reduction to the pressure and temperature at VLT, we obtained
216~mpx/mag which, in limits of errors, agrees to the model (\ref {eq:ladc3})
value.

Use of LADC  considerably reduses the bias caused by atmospheric color effects. 
At precise presetting of LADC  $z _ {\rm L} =z$, 
the bias $S $ is minimized  to $S = d \tan z _ {\rm {L}} $ with coefficient
$d $  20 times smaller in comparison to atmospheric coefficient $ \delta $.

\begin{figure}[htb]
\resizebox{\hsize}{!}{\includegraphics{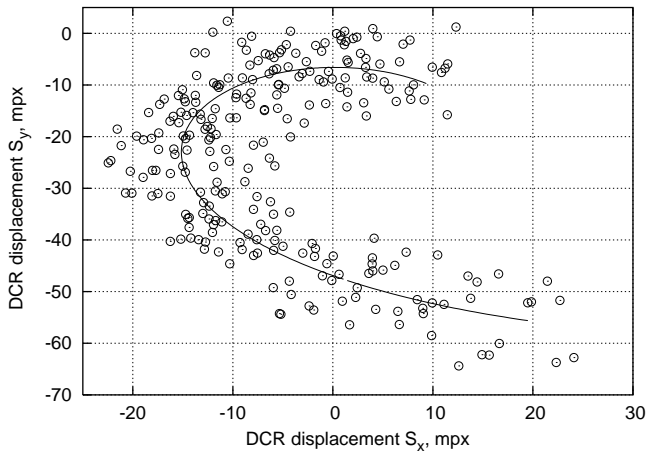}}
\caption {Chromatic image motion  (counter-clockwise and downwards)
of a red $R=17.5$~mag star with $v-r=1.43 $ over a CCD
for 4 hours: points -- measured positions; solid curve --
approximation (\ref {eq:ladc2}) 
}
\label{cxy1115}
\end{figure}

Image displacement induced by DCR effect can be rather considerable.
For example, for a red star with $v-r=1.43 $ and 
local reference field average color $ \overline {v-r} =0.82 $,
the amplitude of the displacement at hour angles $t \sim 3.5 $ 
reaches 60~mpx with an average 20~mpx/hour rate of image
motion (Fig. \ref {cxy1115}). For our session of observations, 
LADC was set at  $z _ {\rm {L}} =45 \degr $ while in a meridian 
$z _ {\rm m} =37.3\degr $.
                          
DCR  effect is a cause of proper motion systematic errors 
proportional to the difference of 
$S_x, S_y $ values  related to the median moments of frames representing 
Ep1 and Ep2.
In view of this, observations should be carried out at a minimum difference
of hour angles $ \Delta t $. Let's estimate restriction on $ \Delta t$ 
if  star  colors and  coefficients of Eqs.(\ref {eq:ladc3}) are known.
For observations in meridian at
$z _ {\rm L} =z _ {\rm m} $, Eq.(\ref {eq:ladc1}) is reduced to $S = d \tan z _
{\rm {m}} $. 
Considering $ \gamma \sim t $, from Eqs.(\ref {eq:ladc2}) one can see that
the largest error
$   \sigma (S_x) \sim \Delta t  \sigma (d) \tan z _ {\rm {m}} $
dependent on  $\Delta t$ is expected in $S_x $.
At the desired  accuracy, say $\sigma (S_x) =0.1$~mpx 
and $\sigma (d) $  quoted above, 
a typical restriction on $ \Delta t $ is
roughly 20~min irrespective of a star color. The precision of model 
(\ref{eq:ladc3}), thus, is marginal to calibrate near meridian observations
obtained with the filter $R _ {\rm special} $.

Restrictions are much stronger for observations at large hour angles.
In this case the main error is caused by the $\delta$ term
in Eq. (\ref {eq:ladc1}) and even with use of colours a good 
calibration of DCR  displacementes is possible only at a very 
strong restriction $ \Delta t \leq 1-5 $ ~ min.

\section{Conclusion}

This study is a first practical test  of an astrometric method based on 
symmetrization of reference fields. We have shown that, in contrast 
to the common belief,
differential astrometric observations with large telescopes (prividing
an adequate reduction performed) are not atmospheric limited. 
At optimal reduction, the photon centroid noise
from reference field is exactly equal to the atmospheric noise. 
Considering extra  cetroiding errors from the target star, 
the photon noise  component {\it always} dominates over the image motion
(Table 3).

This study allows us to predicate that astrometric 
precision of observations at VLT
is 30--50~$\mu$as for stars of 14--16~mag at  10~min exposure (a
session of observations taking about 40 minutes). This precision
is sufficiently  good 
for determination of precise parallaxes and proper motions of
stars, examination of microlensing effects and  exoplanets.
Of course, the measured 
parallaxes and proper motions would be relative. Moreover, 
since each target star position
is measured with reference to its peculiar set of background stars,
zero-points of parallaxes and proper motions are different for
different stars. This circumstance, however, is not critical
for the purposes of exoplanet and (in a minor extent) for
microlensing studies which require only relative astrometric data. 
If necessary, zero-points can be easily reduced to a common system by
iterative procedure.

The most essential restriction for VLT is the saturation of bright images.
Also, sufficiently large light signal necessary for a good referencing
can be obtained if 
observations are restricted to galactic latitudes 20--30$\degr$.
Atmospheric color effects, generally rather large, can be compensated 
by relevant calibrations based on  star colors and
applying restrictions on hour angles. Long-term systematic errors were not
considered in this study and may be a source of additional bias.

Existing filled aperture 8--10~m telescopes are powerful astrometric tools due
to effective averaging of atmospheric fluctuations of a phase over the
aperture and by providing strong light signals. For larger telescopes, in
view of dependences $ \varepsilon
\sim D ^ {-1} $, $ \sigma _ {\rm {rf}} \sim D ^ {-1} $ and 
$ \sigma _ {\rm{at}} \sim D ^ {-3/2} $ 
a further increase of the accuracy is expected . 
Interesting possibilities are  allowed 
by apodization of an entrance
pupil (Paper I) and, in particular, by use of adaptive optics which improves
both centroiding precision and removes low-frequency components of image
motion spectrum.



\begin{acknowledgements}

I would like to thank Prof.~M.Mayor for suggesting to perform this interesting
study, 
Dr.~C.Moutou who kindly provided me with CCD frames and Dr.~L.R. Bedin
for many helpful remarks on different aspects of this Paper.

\end{acknowledgements}

\input ref.tex

\end{document}